\documentclass[numberedappendix,iop]{emulateapj}

\usepackage{graphicx}
\usepackage{natbib}
\usepackage{ulem}
\usepackage{txfonts}

\slugcomment{Submitted to the Astrophysical Journal}

\shorttitle{}
\shortauthors{Mart\' inez Gonz\'alez et al.}

\begin{document}

\title{Spectro-polarimetric Imaging Reveals Helical Magnetic Fields in Solar Prominence Feet}

\author{M. J. Mart\'\i nez Gonz\'alez\altaffilmark{1,2}, R. Manso Sainz\altaffilmark{1,2}, A. Asensio Ramos\altaffilmark{1,2}, 
C. Beck\altaffilmark{1,2,4}, J. de la Cruz Rodr\'\i guez\altaffilmark{3}, A.J. D\'\i az\altaffilmark{1,2,5}}
\altaffiltext{1}{Instituto de Astrof\'\i sica de Canarias, V\'\i a L\'actea s/n, E-38205 La Laguna, Tenerife, Spain}
\altaffiltext{2}{Dept. Astrof\' isica, Universidad de La Laguna, E-38206, La Laguna, Tenerife, Spain}
\altaffiltext{3}{Institute for Solar Physics, Dept. of Astronomy, Stockholm University, Albanova University Center, SE-10691 Stockholm, Sweden}
\altaffiltext{4}{Present address: National Solar Observatory, Sacramento Peak P.O. Box 62, Sunspot, NM 88349, USA}
\altaffiltext{5}{Present address: Departament de F\'\i sica, Universitat de les Illes Balears, E-07071 Palma de Mallorca, Spain}

\begin{abstract}
Solar prominences are clouds of cool plasma levitating 
above the solar surface and insulated from the million-degree corona 
by magnetic fields. 
They form in regions of complex magnetic topology, characterized by 
non-potential fields, which can evolve
abruptly, disintegrating the prominence and ejecting magnetized material into the heliosphere. 
However, their physics is not yet fully understood because
mapping such complex magnetic configurations and their evolution  
is extremely challenging, and must often be guessed by proxy from
photometric observations.
Using state-of-the-art spectro-polarimetric data, 
we reconstruct the structure of the magnetic field in a prominence. 
We find that prominence feet harbor helical 
magnetic fields connecting the prominence to the solar surface below. 
\end{abstract}

\maketitle

\begin{figure*}
\centering
\includegraphics{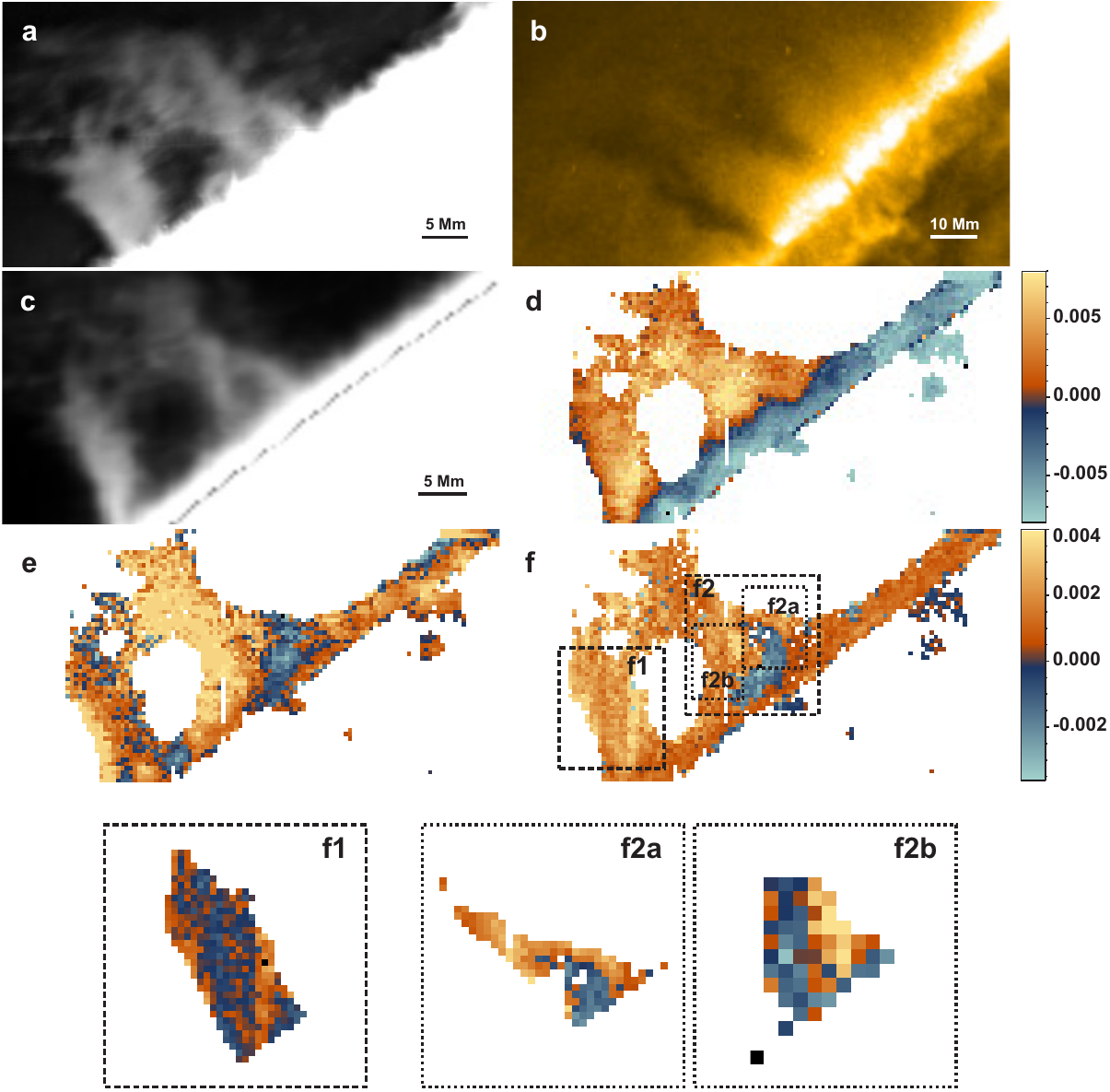}
\caption{{\bfseries a}, Prominence seen at the core of H$_\alpha$. 
This snapshot was recorded half way of the spectropolarimetric scan. 
{\bfseries b}, Prominence as seen at 17.1 nm. 
This snapshot was recorded 10 h after the spectro-polarimetric scan, 
when the pillars of the tornadoes were more prominent 
(no need for sharp-marsking filter) than at the time of the observations. 
{\bfseries c}, Reconstructed intensity map at the core of the He I 1083.0~nm line. 
{\bfseries d, e}, Maps of the amplitude of linear polarization Stokes $Q$ and $U$. 
The reference direction for positive Stokes $Q$ is parallel to the limb. 
{\bfseries f}, Map of the amplitude of circular polarization Stokes $V$. 
Insets show detailed views of the circular polarization in both feet. 
Insets f1-f2b: Stokes $V$ in the region after subtracting the average value (f1 and f2b)}
\end{figure*}

\section{The empirical study of magnetic fields in solar prominences}

Solar prominences are seen as bright translucent clouds at $\sim$10 Mm over the 
solar limb because they mainly scatter light from the underlying disk.
When seen on the solar disk, they appear as dark, long filamentary structures, 
hence called filaments. The main body of the filament (spine) 
often has shorter side-wards extensions ({filament barbs}) which, 
when seen at the limb, give the impression of extending from the spine to the photosphere below
(prominence feet) \citep{mackay_10}. 

It has long been clear that a magnetic field supports the dense material 
of prominences against gravity and prevents them from dissipating into the 
faint, extremely hot corona. Local dips on magnetic field lines can support the plasma, 
and could be induced by the dense, 
heavy prominence plasma itself \citep{kippenhahn_57}, 
or they can exist in force-free \citep{aulanier_98,antiochos_94} 
or stochastic magnetic fields \citep{vanBallegooijen_10}. Yet, all these theoretical claims 
must be constrained by the empirical determination of magnetic fields in 
prominences. 

In the Sun, and in any astrophysical plasma in general, we are not able to directly measure 
these fields, but we are obliged to infer them from the light they emit. Spectro-polarimetry, the 
measurement of the polarized spectrum of light allows us to 
recover {\it quantitative} information on the magnetic field vector. 
The polarization state of observed light is compatible
with the intrinsic (broken) symmetries of the emitting plasma, in particular, 
with the presence of a magnetic field. 
Thus, for example, the emission by an isotropic (and therefore unmagnetized) medium is unpolarized. 
Polarized emission along a magnetic field is
circularly polarized while, normally to the field, it is
linearly polarized, as in the longitudinal or transversal Zeeman effects, 
respectively \cite[e.g.][]{libro_egidio}. 
If light is scattered, additional symmetries are broken and the 
dependencies of polarization are more involved. 
In solar prominences and filaments, spectral lines are polarized by scattering 
and the Zeeman effect, and futher modified by the Hanle effect 
\citep{tandberg_95, libro_egidio}, providing direct information on the magnetic 
field vector. 

Many studies have observed prominences with the aim to determine magnetic fields. 
Some maps of the magnetic field vector in quiescent prominences show 
horizontal magnetic fields of $\sim10-20$ G \citep{casini_05, david_14}, 
in agreement with results obtained in the 1970's and 1980's from 
observations with limited spatial resolution \citep{sahal-brechot77, leroy_89}. 
In contrast, vertical fields have been also diagnosed in prominences \citep{merenda_05}. 
Considering prominence feet, the observation of vertical velocities in these 
structures suggest vertical fields 
directly connecting the spine with the photosphere \citep{zirker_98}. 
Also, from observations of photospheric magnetic fields (not the prominence itself), 
the barbs are interpreted as a series of 
local horizontal dips sustaining plasma at different heights \citep{arturo_06}. 

In the light of these results it is clear that, from the observational point of view, 
the precise topology of magnetic fields in prominences is still a matter of debate. 
The main reasons are that 1) 
measuring the polarized spectrum of solar prominences is an observational challenge, and 
2) the inference of the magnetic field vector in the Hanle regime is subject to 
potential ambiguities. In this paper, we reconstruct the 
topology of the feet of a quiescent prominence from spectro-polarimetric 
data at the He\,{\sc i} 1083.0 nm line. We study the possible solutions 
to the inverse problem and, in contrast to previous results, we discard some of them using a physical constraint. 
We also propose an analytical method to be used to find the multiple solutions in the case of Hanle 
diagnostics in prominences and write the explicit equations in the Appendix.

\section{Near-infrared spectro-polarimetry and multiwavelength imaging}

On April 24 2011 (10:00-13:00 UT), we performed four consecutive spectro-polarimetric 
scans of a quiescent prominence (located at 90E 42S) focusing at the 1083.0~nm multiplet using 
the Tenerife Infrared Polarimeter \citep{tip} at the German Vacuum Tower 
Telescope in the Observatorio del Teide. We integrated 30 s per scan step to reach a polarimetric 
sensitivity of 7$\times$ 10$^{-4}$ times the maximum intensity. Each scan of the prominenece took around 
30 min. This data set constitute a unique time series 
of high polarimetric sensitivity and unprecedented spatial resolution ($\sim 470$ km on the Sun) of a prominence. 
We applied standard reduction procedures to the raw spectra (bias and flat-field correction, 
and polarization demodulation) to obtain maps of the four Stokes parameters I, Q, U, and V (Fig. 1c-f). 
The slit was always kept across the solar limb (horizontal direction in the images), 
which allowed us to correct for seeing-induced cross-talk and stray light \citep{yo_12}. The 
seeing conditions were excellent, the adaptive optics system often reaching an apparent mirror diameter of 
20 cm. This made the applied seeing-induced corrections to be very small, i.e., close to the limb, where the effects 
of seeing are expected to be the largest, the average corrections applied were $5\times$ 10$^{-5}$ times the maximum intensity.

Simultaneous images were taken with a narrow-band Lyot filter centered at the core of the H$_\alpha$ line 
with a cadence of 1 s. These images were treated with blind deconvolution techniques \citep{vanNoort_05} 
to resolve very fine spatial details of the temporal evolution of the prominence 
(Fig. 1a and the online version of the H$_\alpha$ movie). Further context was provided 
by imaging in the coronal line of Fe IX at \hbox{17.1 nm} (Fig. 1b) observed with the Atmospheric 
Imaging Assembly \citep[AIA;][]{aia} instrument onboard the Solar Dynamics Observatory \citep{sdo}. 

\begin{figure}
\center
\includegraphics[width=0.5\columnwidth]{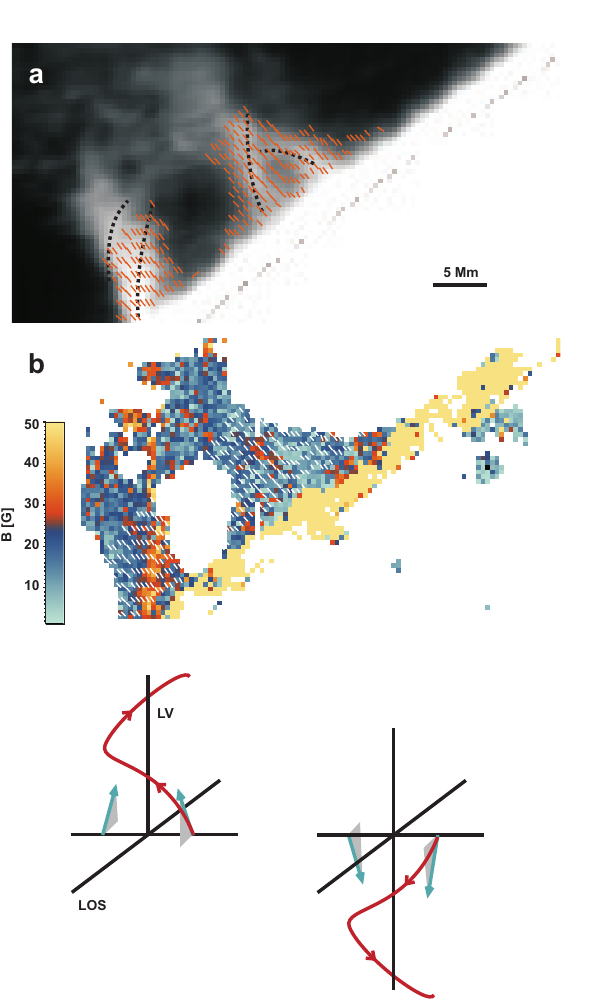}
\caption{{\bfseries a}, Inferred magnetic field topology of the prominence feet.
Direction of the field projected onto the plane of the sky (short lines) 
overplotted to the intensity of the He I 1083.0 nm line and magnetic field strength ({\bfseries b}).
Short-dashed black lines in panel {\bfseries a} trace the axes of the fibrils in the two feet.
In the sketches below, blue arrows represent the actual inferred field vector in f2a fibril. 
Note the opposite polarities of the magnetic field in both sides of the local vertical.
This, and the relative inclination of the projected field with respect to the 
fibrils imply a helical field (red lines).}
\end{figure}

In this paper, we focus on the study of the third scan because of the double-helix appearence 
of the prominence feet. Figure 1 displays the multiwavelength intensity, and polarimetric imaging 
of the observed prominence. Both in the He I 1083.0 nm scan and H$_\alpha$ intensity images, 
one of the two prominence feet (f2 from Fig. 1f) shows a clear double-helix structure formed by two fibrils. 
A more compact, twisted helical structure can also be guessed in foot f1. The two prominence feet 
correspond to vertical, dark (absorption) structures observed at 17.1 nm that resemble those recently 
named {\it solar tornadoes} \citep{su_12, wedemeyer_13}. They were observable in the AIA data for almost 
three more days, and on April 26 (23:38 UT), they suddenly erupted, 
showing a clear helical shape. Interestingly, the foot f2 presents opposite polarities 
of the Stokes $V$ parameter at both sides of one fibril (f2a). This means that the 
magnetic field has reversed polarities along the line of sight at both sides of this fibril. Moreover, 
when subtracting the average circular polarization (i.e., the mean longitudinal magnetic field) to the 
other fibril of feet f2 (panel f2b) and to the other feet (f1), we find similar patterns.

\section{Inference of the magnetic and dynamic structure of the prominence feet}

We analyze the spectro-polarimetric data using the numerical code HaZeL \citep[Hanle Zeeman Light;][]{hazel} 
to recover the full magnetic field vector and the thermodynamical 
properties of the plasma. Facing an inverse problem with observational data and a large number 
of dimensions is always ill-posed. A classical inversion code such as HaZeL retrieves one atmospheric model that 
fits the observed profiles, though others may exist. In our case, the number of these ambiguous solutions depend 
on the regime of the magnetic field and, more importantly, on the scattering geometry. 
Our approach is to find compatible solutions of the same inverse problem
in each pixel and then select the global scenario physically compatible with the context.
This procedure step is essential to reconstruct the global topology of 
the magnetic field in the prominence. 

\subsection{Determination of the scattering geometry}

The angle of the emitting atom in the local vertical (or the observed strcuture if we assume it in the same plane) with respect to the line-of-sight 
(the scattering angle $\theta$) is a very important parameter to correctly infer 
the magnetic field vector from spectro-polarimetric signals generated from 
scattering processes. In order to determine it, 
we used the images of the Atmospheric Imaging Assembly (AIA) at 17.1 nm in which we 
identified the feet of our observed prominence as two dark (absorption) vertical filaments (Fig. 1b). We followed these 
filaments as they entered onto the disk and detected their positions 
(as a projection onto the plane of the sky). We measure the projected distance of the filaments to the limb $d$ over 
time $t$ and fit it with the approximate expression
\begin{equation}
\frac{d}{R} \sim \frac{1}{2}\omega^2 (t-t_{limb})^2,
\label{scatt_angle}
\end{equation}
inferring the value of $t_{limb}$, the time at which the prominence feet were at the limb. The symbol $R$ 
is the radius of the parallel at latitude 
$l=42^\circ$, and $\omega = 0.55^\circ$ h$^{-1}$ 
is the solar angular velocity at that latitude. 
The distance $d$ was obtained as the length of the horizontal 
line from the foot to the limb, which is a good approximation close to the limb. 
The inferred scattering angle (the angle between the line of sight and the local vertical) is approximately given by 
\begin{equation}
\theta \sim -\cos{l}\,{(2d/R)_{obs}}^{1/2}.
\end{equation}
The distance $(d/R)_{obs}$ obtained by substituting $t=t_\mathrm{obs}$, i.e., the time of the observations, 
in Eq. \ref{scatt_angle}. After this procedure, we obtain that the spectro-polarimetric scan was taken at 
a scattering geometry of $\theta=98^\circ$, i.e. while the prominence  was slightly behind the limb.

\begin{figure}
\center
\includegraphics[width=\columnwidth]{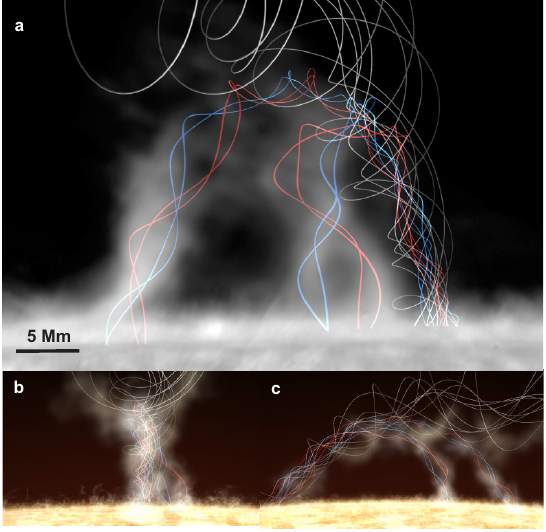}
\caption{Artistic representation of the three-dimensional magnetic field in the 
observed.
In {\bfseries a}, blue and red lines represent arbitrary, recovered field lines 
in the prominence feet; gray lines follow field lines in the spine according 
to overall models for prominence structure and are only drawn as context (i.e. the 
twist of field lines does not represent reality).
Background image: intensity in the He I 1083.0 nm line.
{\bfseries b, c}, Side views (see the online movie of the three-dimensional reconstruction).}
\end{figure}

\subsection{Determination of multiple solutions}

We follow a 2-step inversion scheme to obtain a robust convergence of the code. 
Since the magnetic field is of second order to the intensity 
profile, we first use the intensity profile 
alone to infer the thermodynamical quantities. On a second step, we fix the thermodynamical 
parameters and find the magnetic field vector using the information of the polarization profiles.

As stated before, this may not be the unique solution to the problem. In order to capture all possible 
solutions we follow, again, a 2-step procedure. First, we sample the space of parameters with approximate (though 
appropriate) analytical expressions. Finally, we use these analytical solutions as initial guesses 
for a second HaZeL inversion. This will allow us to refine the 
solutions in the general unsaturated regime to overcome the approximations we have made.

In the case of an optically thin plasma, a normal Zeeman triplet, 
and a magnetic field in the saturated Hanle effect regime 
-- when the Larmor frequency is much larger
than the inverse of the characteristic time for scattering--, the 
geometric dependencies of polarization are simply expressed through 
the Stokes parameters as \citep{casini_05}:
\begin{eqnarray} 
Q&\propto& (3 \cos^2{\theta_B}-1)\sin^2{\Theta_B}\cos{2\Phi_B}\nonumber \\ 
U&\propto& (3 \cos^2{\theta_B}-1)\sin^2{\Theta_B}\sin{2\Phi_B}\nonumber \\ 
V&\propto& \cos{\Theta_B},
\label{eqs}
\end{eqnarray}
showing a dependence on the inclination $\Theta_B$ and azimuth $\Phi_B$ 
of the magnetic field with respect to the line-of-sight (LOS), like in the 
Zeeman effect, and an additional dependence on the geometry 
of the scattering event through the inclination $\theta_B$ of the magnetic field 
with respect to the local vertical (LV). 
The linearly polarized components (Stokes $Q$ and $U$) are dominated by scattering 
polarization and the Hanle effect, while the circularly polarized component (Stokes $V$) 
is generated by the longitudinal 
Zeeman effect, and defines the polarity of the magnetic field along the LOS \citep{libro_egidio}. 

Equations~\ref{eqs} show a dependence of polarization on $\Theta_B$ and $\Phi_B$ 
that yield the well-known $180^\circ$-ambiguity for $\Phi_B$ of classical Zeeman diagnostics. 
The additional dependence on $\theta_B$, characteristic of scattering processes,
may yield two additional ambiguous solutions for the magnetic field \citep{casini_05}.
The He~I 1083.0 nm line often forms close to the saturation regime 
described by Eqs. \ref{eqs} and all (up to four) ambiguous field configurations 
can be determined analytically (see the Appendix) \citep[see also][]{judge_07}. 

\begin{figure}
\center
\includegraphics[width=\columnwidth]{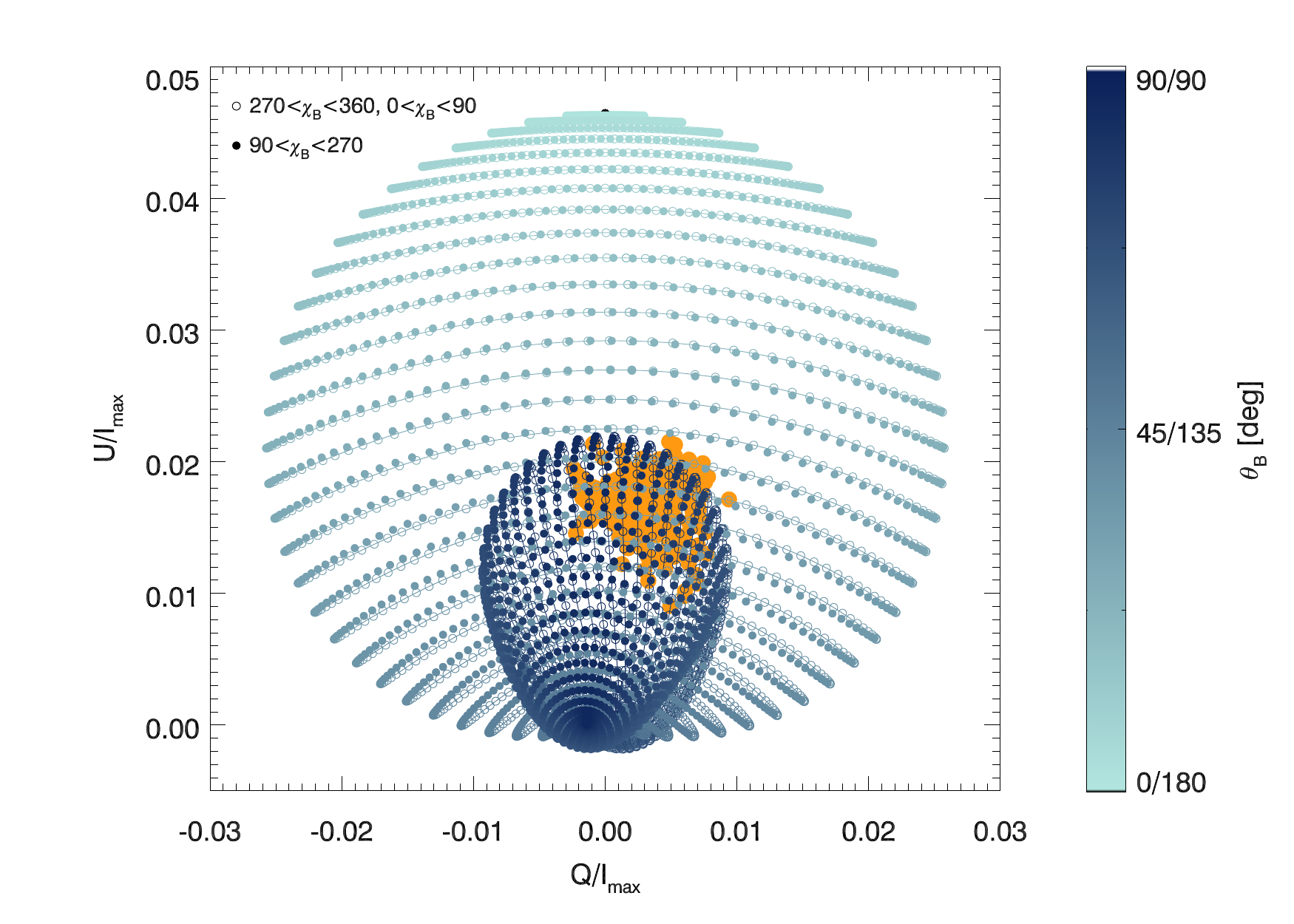}
\caption{Hanle effect diagram for the He\,{\sc I} 1083.0 nm line computed 
with the HaZeL code assuming a 98$^\circ$ scattering geometry and the saturation regime (B=30 G). The 
inclination ($\theta_B$) and azimuth ($\phi_B$) are referred to the LV. The $\chi_B=0$ corresponds to the line 
of sight, and $\chi_B=0$ for a radial vector away from the Sun \citep[see]{hazel}. The solid 
line represent the curves at constant inclination. The Hanle diagram for the range of inclinations 
between 90$^\circ$ and $180^\circ$ is the same but with a reversed color palette. The orange dots represent 
the observed polarization amplitudes.All the observed points have ambiguous vertical and horizontal solutions. 
However, note that the sign of the circular polarization is only compatible with 
one of the solutions represented by filled or empty dots.}
\label{diagrama}
\end{figure}

In order to have an idea of the potential multiple solutions we face, we have 
represented in Figure \ref{diagrama} the so-called Hanle diagram: the amplitude of Stokes $Q$ and $U$ 
in terms of the inclination ($\theta_B$) and azimuth ($\phi_B$) of the magnetic field in the LV. 
The diagram is computed for a point at 15$''$ above the solar limb (middle-upper parts of the observed 
prominence feet). We have assumed the saturation 
regime (in particular a field strength of 30 G), and the scattering angle determined by our observations (98$^\circ$). 
The zero inclination is defined as a field in the LV, and the zero azimuth is defined in the LOS, being $\phi_B=180^\circ$ 
a field directed away from the observer. 

The more vertical fields ($\theta_B \le 35^\circ$ and $\theta_B \ge 145^\circ$) have four possible solutions irrespective
of the azimuth value: two of them (the 
ones represented by filled and empty circles) are not ambiguous if circular polarization is observed, since it 
provides the sign of the longitudinal component of the magnetic field. The other two solutions have opposite polarities of the 
field in the LV. Fields with inclinations between $35^\circ$ and $145^\circ$ have eight possible solutions: 
a more vertical inclination with the four solutions stated above and another four solutions 
with more inclined fields. 
In the case of our observations, the sensitivity of circular polarization allows us to constrain the azimuth range, 
and hence only four ambiguous solutions need to be determined. The polarity of the field (in the LV) 
can not be determined. However, as we will see, this ambiguity is unimportant for the purposes of this paper. 

\begin{figure}
\includegraphics[width=\columnwidth]{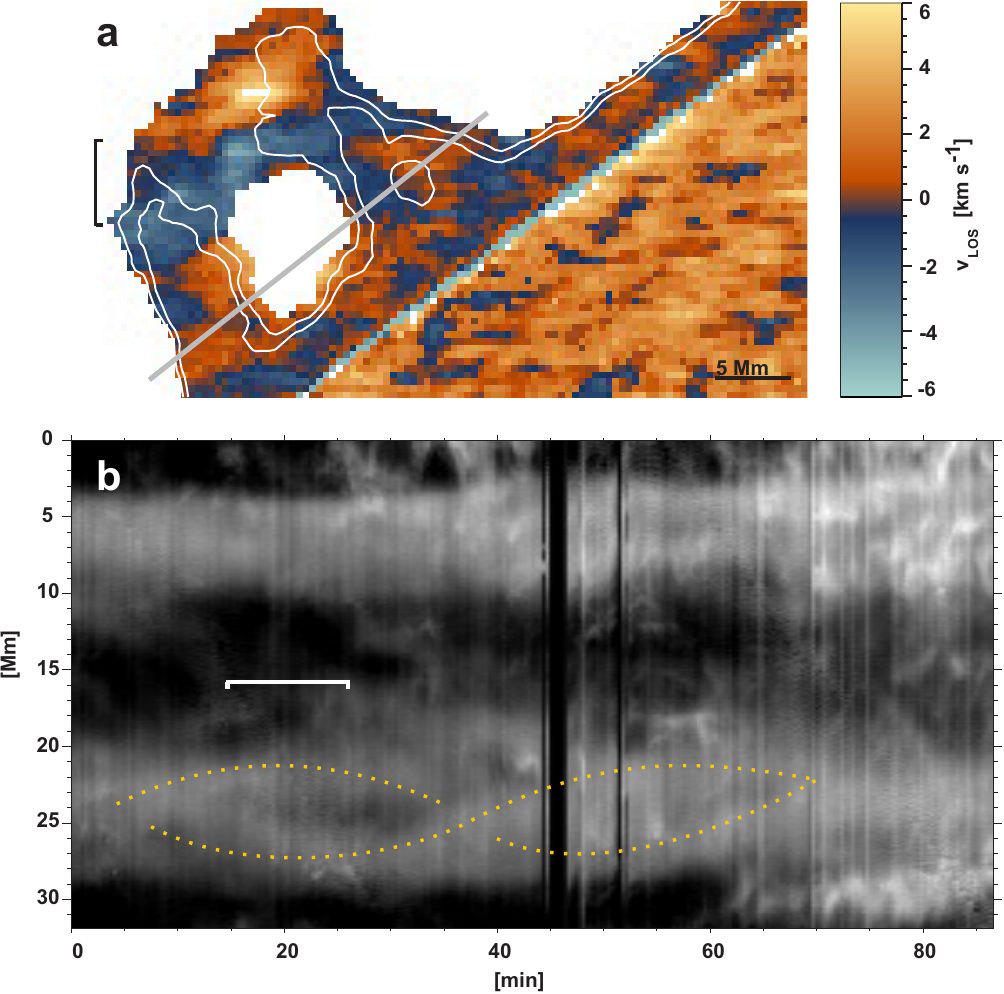}
\caption{{\bfseries a}, LOS velocity field from the inversion of the He I 1083.0 nm observations 
with intensity contours (white lines) overplotted for reference. 
{\bfseries b}, Space-time variations of H$_\alpha$ intensity for an artificial slit 
at 5.8 Mm, parallel to the limb across the helical structure (gray line). 
Brackets mark the time interval during the scan of f2. 
Dotted yellow lines trace the periodic movement of the two fibrils in the double-helix. 
The composition of both motions indicate rotation of the structure.}
\end{figure}

\subsection{The helical magnetic field}

The ambiguities apply at each pixel of our observations. However, we assume that there are no 
tangential discontinuities or shocks and hence the 
magnetic field in the prominence draw continuous lines. We reconstructed four global topologies 
of the magnetic field 
which could be grouped into two broad categories: one with fields inclined by 
$\theta_B \sim 30^\circ, 150^\circ$, and another one with more inclined fields ($\theta_B \sim 90^\circ$). 
In both cases, the projection of the field onto the plane of the 
sky is at an angle with the axis of each fibril that form the prominence feet.
The fibril f2a (displayed in Fig. 1) shows a LOS polarity reversal at opposite sides of its axis. 
Similar LOS polarity inversions appear across the axes of the other fibrils 
when the average value of Stokes V in the region, corresponding to the mean
LOS component of the magnetic field, is subtracted (Figs. 1f1 and 1f2b). 
Put together, this points to an helical global topology in both families of solutions,
with the more horizontal configurations showing a more twisted field.

In order to disambiguate the problem, we introduce a physical constraint through a 
stability analysis. According to the Kruskal-Shafranov criterion \citep{hood_79} 
a kink instability develops when the amount of magnetic twist exceeds 
a critical value, so that a structure is stable if
\begin{equation}
\frac{2\pi r}{L}\frac{B_z}{B_\theta} \ge 1,
\end{equation}
where $r$ and $L$ are the radius and the length of the fibril, respectively, 
and $B_z$ and $B_\theta$ are the vertical and azimuthal components of the magnetic field, 
respectively. 
Estimating \hbox{$r=1.7$ Mm} and \hbox{$L= 11.6$ Mm} for our prominence, the stability criterion 
yields 0.09 for the solution with the fields around 90$^\circ$ and 1.28 for the fields with inclinations 
of $30^\circ, 150^\circ$.

Figure 2 shows the magnetic field strength and topology of 
the magnetic field. The magnetic field strength has very similar values 
throughout all families of solutions. In most of the prominence the 
magnetic field strength is below 20~G, but shows 
filamentary structures, parallel to the helix structure, 
of higher field strength ($\sim 60$ G; Fig. 2b). 
The projection of the field onto the plane of the sky is always roughly perpendicular
to the solar limb, and at an angle ($\sim 30^\circ$-$50^\circ$) with 
the axis of the fibrils that form the prominence feet (Fig.~2a).
All fibrils present magnetic field polarity reversals at both sides 
of their axis. This implies a helical field along the fibrils with the axis in the plane 
of the sky in f2a (since the average longitudinal magnetic field is close to zero), and slightly 
tilted relative to the LV in the other fibrils (since polarity reversals are only seen when the mean 
longitudinal magnetic field is subtracted). 
The helical magnetic fields of the two fibrils are interlaced to form a double-helix 
that constitutes one foot of the prominence (see Fig. 3 and the online movie of the artistic 
representation of the field topology). The feet magnetically connect the spine with lower layers, 
in contrast to previous works suggesting that feet 
are just a collection of dips at different heights \cite{arturo_06}. 

\subsection{Motions of the prominence material}

The intensity profiles of the He I line carry information on the LOS velocity, 
opacity, temperature, and density of the structure. We recover all these parameters along with 
the magnetic topology using the HaZeL code. They have the same values for all families of 
solutions of the magnetic field. 

Figure 5a shows the inferred LOS velocities. Interestingly, they have opposite signs at both sides 
of both feet of the prominence (f1 and f2), with values up to \hbox{2 km s$^{-1}$} 
\cite[similar values as][]{david_12}. Assuming that the neutral atoms of He still trace the field lines, 
this pattern of velocities could be a natural 
consequence if the prominence material is flowing along the inferred double-helix structure of the prominence feet. 
Note that f1 has a positive-negative 
Doppler pattern at lower heights, while it has a negative-positive one higher in the feet, as 
we would expect for a more compact double-helix, which implies a larger twist.

If the material is flowing along the helical field lines of the fibrils of each feet, a Doppler 
pattern similar to that of large scales should be observed at smaller scales. The upper parts of f2a show a positive-negative pattern. The rest 
of the fibrils have a continuous increase of the Doppler velocity as we move from left to right across the structure. 
This could be interpreted as a negative-positive pattern if the fibrils are not at a 90$^\circ$ scattering geometry, which is very likely. 

The two fibrils of f2 exhibit a periodic motion in the H$_\alpha$ space-time diagram, with a period of 
$\sim 60$ min (Fig. 5b), consistent with the periods reported in \cite{su_12} for solar tornadoes. 
Assuming that the change in H$_\alpha$ brightness is only due to plasma movements, the 
observed periodic motion could be also interpreted in terms of plasma flowing along helical field lines. 
Rotation of the magnetic structure is very unlikely since the observed $\sim 50$ min period 
would imply a full turn of the magnetic field in less than one hour. This will 
considerably increase the twist of the structure and will soon make it unstable under magnetohydrodynamical 
instabilities (easily within a day).
The plane-of-the-sky periodic motions show maximum tangential velocities of $\sim 9$ km s$^{-1}$, 
which are far larger than the ones inferred from the Doppler effect in the He\,{\sc i} line. 
We must be careful to directly relate those two velocities since Doppler measurements are generated only by plasma motions 
while changes in H$_\alpha$ brightness can not only be assigned to plasma movements but to changes in the thermal conditions.

\section{Discussion}

The feet of the observed prominence harbor helical magnetic fields. After assuming a 
simple stability criterion, we have a preference for the more vertical solution and conclude 
that the magnetic fields in prominence feet 
connect the spine with the underlying atmosphere. These results are in contrast to the scenario in 
which these structures are formed by a series of 
local horizontal dips that sustain the plasma at different heights \citep{arturo_06, aulanier_99}. 
Assuming a uniformly twisted straight cylinder, the 
magnetic field displays a twist (number of turns over its length $L$) of 1.32 at each fibril. 
We speculate that the connectivity of the prominence spine, 
with a well-defined helicity, and the photospheric magnetic field below, 
with a fluctuating topology, may naturally yield the kind of helical structures we find. 

The He\,{\sc i} Doppler velocities display opposite velocities along the LOS at both sides of 
prominence feet. The H$_\alpha$ intensity displays periodic motions in the plane-of-the-sky. 
Using only the Doppler velocities or the plane-of-the-sky motions alone 
do not allow to reach any conclusion on the plasma motions and can lead to controversies in 
the literature \citep[e.g.][]{panasenco_14, david_12}. If we put the 
He\,{\sc i} and the H$_\alpha$ information together, we could interpret these observations 
as the material of the prominence flowing along helical field lines. However, the LOS Doppler 
velocities inferred from the He\,{\sc i} line and the tangential velocities expected for the 
observed period of the H$_\alpha$ images do not match. It could be that 
neutral H and He clouds have different thermodynamical properties or it could be that 
the changes in the H$_\alpha$ brightness have an important contribution from changing thermal conditions,
but we need more data to really understand this issue. 

The prominence remains stable for several days, and we think that the magnetic topology 
of the prominence feet reported in this paper plays a fundamental role on the stability 
of the prominence as a whole. 

\begin{acknowledgements}
The authors are especially grateful to Eric Priest, Manuel Collados, and I\~nigo Arregui for very interesting 
discussions on solar prominences, and for 
their enthusiasm on this work. We are also grateful to Arturo L\'opez Ariste, Manuel Luna, David Orozco Su\'arez, 
and Javier Trujillo Bueno for very helpful discussions. This work is based on observations made with the Vacuum Tower 
Telescope 
operated on Tenerife by the Kiepenheuer-Institut f\"ur Sonnenphysik in the Spanish Observatorio del Teide of the 
Instituto de 
Astrof\'\i sica de Canarias. Financial support by the Spanish Ministry of Economy and Competitiveness and the 
European FEDER Fund through project AYA2010-18029 (Solar Magnetism and Astrophysical Spectropolarimetry) is 
gratefully acknowledged. AAR acknowledges financial support through the Ram\'on y Cajal fellowship. 
RMS and AAR also acknowledge financial support from the Consolider-Ingenio 2010 CSD2009-00038 project.
\end{acknowledgements}


\begin{thebibliography}{30}
\expandafter\ifx\csname natexlab\endcsname\relax\def\natexlab#1{#1}\fi

\bibitem[{{Antiochos} {et~al.}(1994){Antiochos}, {Dahlburg}, \&
  {Klimchuk}}]{antiochos_94}
{Antiochos}, S.~K., {Dahlburg}, R.~B., \& {Klimchuk}, J.~A. 1994, \apjl, 420,
  L41

\bibitem[{{Asensio Ramos} {et~al.}(2008){Asensio Ramos}, {Trujillo Bueno}, \&
  {Landi Degl'Innocenti}}]{hazel}
{Asensio Ramos}, A., {Trujillo Bueno}, J., \& {Landi Degl'Innocenti}, E. 2008,
  \apj, 683, 542

\bibitem[{{Aulanier} {et~al.}(1999)}]{aulanier_99}
{Aulanier}, G., {Demoulin}, P., {Mein}, N., {van Driel-Gesztelyi}, L., {Mein}, P., \& {Schmieder}, B. 1999, \aap, 342, 867

\bibitem[{{Aulanier} \& {Demoulin}(1998)}]{aulanier_98}
{Aulanier}, G., \& {Demoulin}, P. 1998, \aap, 329, 1125

\bibitem[{{Casini} {et~al.}(2005){Casini}, {Bevilacqua}, \& {L{\'o}pez
  Ariste}}]{casini_05}
{Casini}, R., {Bevilacqua}, R., \& {L{\'o}pez Ariste}, A. 2005, \apj, 622, 1265

\bibitem[{{Casini} {et~al.}(2003){Casini}, {L{\'o}pez Ariste}, {Tomczyk}, \&
  {Lites}}]{casini_03}
{Casini}, R., {L{\'o}pez Ariste}, A., {Tomczyk}, S., \& {Lites}, B.~W. 2003,
  \apjl, 598, L67

\bibitem[{{Collados}(1999)}]{tip}
{Collados}, M. 1999, in Astronomical Society of the Pacific Conference Series,
  Vol. 184, Third Advances in Solar Physics Euroconference: Magnetic Fields and
  Oscillations, ed. B.~{Schmieder}, A.~{Hofmann}, \& J.~{Staude}, 3--22

\bibitem[{{Hood} \& {Priest}(1979)}]{hood_79}
{Hood}, A.~W., \& {Priest}, E.~R. 1979, \solphys, 64, 303

\bibitem[{{Judge} (2007)}]{judge_07}
{Judge}, P. 2007, \apj, 662, 677

\bibitem[{{Kippenhahn} \& {Schl{\"u}ter}(1957)}]{kippenhahn_57}
{Kippenhahn}, R., \& {Schl{\"u}ter}, A. 1957, \zap, 43, 36

\bibitem[{{Landi Degl'Innocenti} \& {Landolfi}(2004)}]{libro_egidio}
{Landi Degl'Innocenti}, E., \& {Landolfi}, M. 2004, Polarization in Spectral
  Lines (Kluwer Academic Publishers)

\bibitem[{{Lemen} {et~al.}(2012){Lemen}, {Title}, {Akin}, {Boerner}, {Chou},
  {Drake}, {Duncan}, {Edwards}, {Friedlaender}, {Heyman}, {Hurlburt}, {Katz},
  {Kushner}, {Levay}, {Lindgren}, {Mathur}, {McFeaters}, {Mitchell}, {Rehse},
  {Schrijver}, {Springer}, {Stern}, {Tarbell}, {Wuelser}, {Wolfson}, {Yanari},
  {Bookbinder}, {Cheimets}, {Caldwell}, {Deluca}, {Gates}, {Golub}, {Park},
  {Podgorski}, {Bush}, {Scherrer}, {Gummin}, {Smith}, {Auker}, {Jerram},
  {Pool}, {Soufli}, {Windt}, {Beardsley}, {Clapp}, {Lang}, \& {Waltham}}]{aia}
{Lemen}, J.~R., {Title}, A.~M., {Akin}, D.~J., {Boerner}, P.~F., {Chou}, C.,
  {Drake}, J.~F., {Duncan}, D.~W., {Edwards}, C.~G., {Friedlaender}, F.~M.,
  {Heyman}, G.~F., {Hurlburt}, N.~E., {Katz}, N.~L., {Kushner}, G.~D., {Levay},
  M., {Lindgren}, R.~W., {Mathur}, D.~P., {McFeaters}, E.~L., {Mitchell}, S.,
  {Rehse}, R.~A., {Schrijver}, C.~J., {Springer}, L.~A., {Stern}, R.~A.,
  {Tarbell}, T.~D., {Wuelser}, J.-P., {Wolfson}, C.~J., {Yanari}, C.,
  {Bookbinder}, J.~A., {Cheimets}, P.~N., {Caldwell}, D., {Deluca}, E.~E.,
  {Gates}, R., {Golub}, L., {Park}, S., {Podgorski}, W.~A., {Bush}, R.~I.,
  {Scherrer}, P.~H., {Gummin}, M.~A., {Smith}, P., {Auker}, G., {Jerram}, P.,
  {Pool}, P., {Soufli}, R., {Windt}, D.~L., {Beardsley}, S., {Clapp}, M.,
  {Lang}, J., \& {Waltham}, N. 2012, \solphys, 275, 17

\bibitem[{{Leroy}(1989)}]{leroy_89}
{Leroy}, J.~L. 1989, in Astrophysics and Space Science Library, Vol. 150,
  Dynamics and Structure of Quiescent Solar Prominences, ed. E.~R. {Priest},
  77--113

\bibitem[{{L{\'o}pez Ariste} {et~al.}(2006){L{\'o}pez Ariste}, {Aulanier},
  {Schmieder}, \& {Sainz Dalda}}]{arturo_06}
{L{\'o}pez Ariste}, A., {Aulanier}, G., {Schmieder}, B., \& {Sainz Dalda}, A.
  2006, \aap, 456, 725

\bibitem[{{Mackay} {et~al.}(2010){Mackay}, {Karpen}, {Ballester}, {Schmieder},
  \& {Aulanier}}]{mackay_10}
{Mackay}, D.~H., {Karpen}, J.~T., {Ballester}, J.~L., {Schmieder}, B., \&
  {Aulanier}, G. 2010, \ssr, 151, 333

\bibitem[{{Mart{\'{\i}}nez Gonz{\'a}lez} {et~al.}(2012){Mart{\'{\i}}nez
  Gonz{\'a}lez}, {Asensio Ramos}, {Manso Sainz}, {Beck}, \& {Belluzzi}}]{yo_12}
{Mart{\'{\i}}nez Gonz{\'a}lez}, M.~J., {Asensio Ramos}, A., {Manso Sainz}, R.,
  {Beck}, C., \& {Belluzzi}, L. 2012, \apj, 759, 16

\bibitem[{{Merenda} {et~al.}(2006){Merenda}, {Trujillo Bueno}, {Landi
  Degl'Innocenti}, \& {Collados}}]{merenda_05}
{Merenda}, L., {Trujillo Bueno}, J., {Landi Degl'Innocenti}, E., \& {Collados},
  M. 2006, \apj, 642, 554

\bibitem[{{Orozco Su{\'a}rez} {et~al.}(2012){Orozco Su{\'a}rez}, {Asensio
  Ramos}, \& {Trujillo Bueno}}]{david_12}
{Orozco Su{\'a}rez}, D., {Asensio Ramos}, A., \& {Trujillo Bueno}, J. 2012,
  \apjl, 761, L25

\bibitem[{{Orozco Su{\'a}rez} {et~al.}(2014){Orozco Su{\'a}rez}, {Asensio
  Ramos}, \& {Trujillo Bueno}}]{david_14}
{Orozco Su{\'a}rez}, D., {Asensio Ramos}, A., \& {Trujillo Bueno}, J. 2014,
  \apj, 566, 46

\bibitem[{{Panasenco} {et~al.}(2014){Panasenco}, {Martin}, \&
  {Velli}}]{panasenco_14}
{Panasenco}, O., {Martin}, S.~F., \& {Velli}, M. 2014, \solphys, 289, 603

\bibitem[{{Pesnell} {et~al.}(2012){Pesnell}, {Thompson}, \& {Chamberlin}}]{sdo}
{Pesnell}, W.~D., {Thompson}, B.~J., \& {Chamberlin}, P.~C. 2012, \solphys,
  275, 3

\bibitem[{{Sahal-Brechot} {et~al.}(1977){Sahal-Brechot}, {Bommier}, \&
  {Leroy}}]{sahal-brechot77}
{Sahal-Brechot}, S., {Bommier}, V., \& {Leroy}, J.~L. 1977, \aap, 59, 223

\bibitem[{{Su} {et~al.}(2012){Su}, {Wang}, {Veronig}, {Temmer}, \&
  {Gan}}]{su_12}
{Su}, Y., {Wang}, T., {Veronig}, A., {Temmer}, M., \& {Gan}, W. 2012, \apjl,
  756, L41

\bibitem[{{Tandberg-Hanssen}(1995)}]{tandberg_95}
{Tandberg-Hanssen}, E., ed. 1995, Astrophysics and Space Science Library, Vol.
  199, {The nature of solar prominences}

\bibitem[{{van Ballegooijen} \& {Cranmer}(2010)}]{vanBallegooijen_10}
{van Ballegooijen}, A.~A., \& {Cranmer}, S.~R. 2010, \apj, 711, 164

\bibitem[{{van Noort} {et~al.}(2005){van Noort}, {Rouppe van der Voort}, \&
  {L{\"o}fdahl}}]{vanNoort_05}
{van Noort}, M., {Rouppe van der Voort}, L., \& {L{\"o}fdahl}, M.~G. 2005,
  \solphys, 228, 191

\bibitem[{{Wedemeyer} {et~al.}(2013){Wedemeyer}, {Scullion}, {Rouppe van der
  Voort}, {Bosnjak}, \& {Antolin}}]{wedemeyer_13}
{Wedemeyer}, S., {Scullion}, E., {Rouppe van der Voort}, L., {Bosnjak}, A., \&
  {Antolin}, P. 2013, \apj, 774, 123

\bibitem[{{Zirker} {et~al.}(1998){Zirker}, {Engvold}, \& {Martin}}]{zirker_98}
{Zirker}, J.~B., {Engvold}, O., \& {Martin}, S.~F. 1998, \nat, 396, 440

\end{thebibliography}

\appendix

\section{Ambiguities in the Hanle effect in the saturation regime}
In the saturation regime of the Hanle effect, Stokes $Q$ and $U$ are insensitive to the field
strength, but are sensitive to the geometry of the field. For a two-level atom with a $J_{up}=0 \to J_{low}=1$ 
transition, the optically thin limit, and the saturation regime for the Hanle effect, 
the linear polarization can be written as:
\begin{eqnarray}
Q &=& \frac{q}{2} \left( 3 \cos^2 \theta_B-1 \right) \sin^2\Theta_B \cos 2\Phi_B \nonumber \\
U &=& \frac{q}{2} \left( 3 \cos^2 \theta_B-1 \right) \sin^2\Theta_B \sin 2\Phi_B.
\end{eqnarray}
The inclination and azimuth of the magnetic field in the LV are represented by the symbols $\Theta_B$ and $\Phi_B$, 
respectively. In the LOS, the inclination and azimuth are displayed as $\theta_B$ and $\phi_B$. These equations 
are formally the same irrespective of the scattering angle $\theta$. The dependence on the scattering geometry is 
implicit in the amplitude in the abscence of a magnetic field $q$. 

The coordinates of the magnetic field vector $\mathbf{B}$ in the reference system of the vertical
and the reference system of the LOS are:
\begin{eqnarray}
\mathbf{B} &=& B \left(\sin \theta_B \cos \phi_B \mathbf{i}+\sin \theta_B \sin \phi_B \mathbf{j}+\cos \theta_B \mathbf{k} \right) \nonumber \\
\mathbf{B} &=& B \left(\sin \Theta_B \cos \Phi_B \mathbf{i}'+\sin \Theta_B \sin \Phi_B \mathbf{j}'+\cos \Theta_B \mathbf{k}' \right),
\end{eqnarray}
where the unit vectors are related by a simple rotation:
\begin{eqnarray}
\mathbf{i}' &=& \cos \theta \mathbf{i} - \sin \theta \mathbf{k} \nonumber \\
\mathbf{k}' &=& \sin \theta \mathbf{i} + \cos \theta \mathbf{k}.
\end{eqnarray}
Given that the magnetic field vector must be the same in both reference systems, we find that the
following relations apply:
\begin{eqnarray}
\sin \theta_B \cos \phi_B &=& \sin \Theta_B \cos \Phi_B \cos \theta + \cos \Theta_B + \sin \theta \nonumber \\
\sin \theta_B \sin \phi_B &=& \sin \Theta_B \sin \Phi_B \nonumber \\
\cos \theta_B &=& \cos \Theta_B \cos \theta - \sin \Theta_B \cos \Phi_B \sin \theta.
\end{eqnarray}
Solving the previous three equations in the two directions, we find the following transformations
between the angles in the vertical reference system and the LOS reference system:
\begin{eqnarray}
\cos \Theta_B &=& \cos\theta \cos\theta_B + \sin\theta \sin\theta_B \cos\phi_B \nonumber \\
\sin \Theta_B &=& +\sqrt{1-\cos^2\Theta_B} \nonumber \\
\cos \Phi_B &=& \frac{\cos\theta \sin\theta_B \cos\phi_B - \cos\theta_B \sin\theta}{\sin \Theta_B} \nonumber \\
\sin \Phi_B &=& \frac{\sin\theta_B \sin\phi_B}{\sin\Theta_B}
\end{eqnarray}
and
\begin{eqnarray}
\cos \theta_B &=& \cos\theta \cos\Theta_B - \sin\theta \sin\Theta_B \cos\Phi_B \nonumber \\
\sin \theta_B &=& +\sqrt{1-\cos^2\theta_B} \nonumber \\
\cos \phi_B &=& \frac{\cos\theta \sin\Theta_B \cos\Phi_B + \cos\Theta_B \sin\theta}{\sin \theta_B} \nonumber \\
\sin \phi_B &=& \frac{\sin\Theta_B \sin\Phi_B}{\sin\theta_B}.
\end{eqnarray}
Note that, since $\Theta_B \in [0,\pi]$, we can safely use the square root and take the positive value.
In order to transform from one reference system to the other, we can compute the
inclination easily by inverting the sinus or the cosinus. However, the situation is different
for the azimuth, because the range of variation is $[-\pi,\pi]$. Therefore, one has to compute the 
cosinus and the sinus separately and the decide which is the correct quadrant fo the angle in terms
of the signs of both quantities.

\begin{center}
\end{center}

Four multiple solutions can exist for the Stokes $Q$ and $U$ parameters. The idea is that $\Phi_B$
can be modified and still obtain the same $Q$ and $U$ by properly adjusting the value of $\Theta_B$. It is
clear that, given that the term that can be used to compensate for the change in the azimuth on the LOS
reference system is the same for Stokes $Q$ and $U$, we can only compensate for changes in the sign. Therefore,
we have the following potential ambiguities:
\begin{eqnarray}
\Phi_B' &=& \Phi_B \nonumber \\
\Phi_B' &=& \Phi_B -\pi/2 \nonumber \\
\Phi_B' &=& \Phi_B + \pi/2 \nonumber \\
\Phi_B' &=& \Phi_B + \pi.
\end{eqnarray}

We have to compute the value of $\Theta_B'$ that keeps the value of $Q$ and $U$ unchanged. Therefore,
once we find a solution to the inversion problem in the form of the pair $(\theta_B,\phi_B)$, we can find
the remaining solutions in the saturation regime following the recipes that we present now. Remember that, unless
one knows the sign of $\cos\Theta_B$ (given by the observation of circular polarization), the number of potential
ambiguous solutions is 8. If the polarity of the field is known, the number is typically reduced to 4 (or 2
if no 90$^\circ$ ambiguity is present).

\subsection{$\Phi_B' = \Phi_B$}
Under this change, we have that
\begin{eqnarray}
\cos 2\Phi_B' &=& \cos 2\Phi_B, \\ \nonumber 
\sin 2\Phi_B' &=& \sin 2\Phi_B, \\ \nonumber
\cos \Phi_B' &=& \cos \Phi_B, \\ \nonumber
\sin \Phi_B' &=& \sin \Phi_B.
\end{eqnarray}
Making use of the previous relations between the angles wrt to the vertical and the LOS, we have to solve the 
following equation:
\begin{equation}
\left( 3 \cos^2\theta_B'-1 \right) \sin^2 \Theta_B' = \left( 3 \cos^2\theta_B-1 \right) \sin^2 \Theta_B,
\end{equation}
which can be written as:
\begin{eqnarray}
&&\left[ 3 \left( \cos \Theta_B' \cos \theta - \sin\theta \sin\Theta_B' \cos\Phi_B\right)^2-1 \right] \sin^2 \Theta_B' \\ \nonumber
&=&\left[ 3 \left( \cos \Theta_B \cos \theta - \sin\theta \sin\Theta_B \cos\Phi_B\right)^2-1 \right] \sin^2 \Theta_B.
\end{eqnarray}
After some algebra and doing the substitution $t=\sin\Theta_B'$, we end up with the following equation to be
solved:
\begin{equation}
A t^4 + Bt^2 + C t^3 \sqrt{1-t^2} = K,
\end{equation}
where
\begin{eqnarray}
A &=& -3\cos^2 \theta + 3\sin^2 \theta \cos^2 \Phi_B \nonumber \\
B &=& 3\cos^2 \theta - 1 \nonumber \\
C &=& -6 \cos\theta \sin\theta \cos \Phi_B \nonumber \\
K &=& \left[ 3 \left( \cos \Theta_B \cos \theta - \sin\theta \sin\Theta_B \cos\Phi_B\right)^2-1 \right] \sin^2 \Theta_B.
\end{eqnarray}
The previous equation can be solved if we make the change of variables $t=\pm \sqrt{Z}$, resulting in:
\begin{eqnarray}
&&(C^2+A^2) Z^4 + (-C^2+2AB) Z^3 + (-2AK+B^2) Z^2 \\ \nonumber 
&-& 2BKZ + K^2 = 0.
\end{eqnarray}
This polynomial of 4-th order can have four different solutions. From these solutions, we have to take only
the real solutions which are larger than 0, given the range of variation of $\Theta_B$:
\begin{equation}
t \in \mathbb{R}, \qquad 0 \leq t \leq 1.
\end{equation}
Once the solutions for $t$ are found, we make $\Theta_B' = \arcsin t$. Note that, for a fixed value of $t$,
two values of $\Theta_B'$ are possible. We choose the correct one by evaluating the expressions for 
$Q$ and $U$ and testing which of the two possible choices give the values equal (or very similar) to the original ones.

The angles $(\theta_B,\phi_B)$ are obtained by doing the transformation from $(\Theta_B',\Phi_B)$ to the
vertical reference system.

\subsection{$\Phi_B' = \Phi_B+\pi$}

Under this change, we have:
\begin{eqnarray}
\cos 2\Phi_B' &=& \cos 2\Phi_B, \\ \nonumber
\sin 2\Phi_B' &=& \sin 2\Phi_B,  \\ \nonumber
\cos \Phi_B' &=& -\cos \Phi_B,  \\ \nonumber
\sin \Phi_B' &=& -\sin \Phi_B.
\end{eqnarray}
Following the same approach, we have to solve for $\Theta_B'$ in 
\begin{eqnarray}
&&\left[ 3 \left( \cos \Theta_B' \cos \theta + \sin\theta \sin\Theta_B' \cos\Phi_B\right)^2-1 \right] \sin^2 \Theta_B' \\ \nonumber 
&=&\left[ 3 \left( \cos \Theta_B \cos \theta - \sin\theta \sin\Theta_B \cos\Phi_B\right)^2-1 \right] \sin^2 \Theta_B.
\end{eqnarray}
The solution are obtained as the roots of the same equations as before but now
\begin{eqnarray}
A &=& -3\cos^2 \theta + 3\sin^2 \theta \cos^2 \Phi_B \nonumber \\
B &=& 3\cos^2 \theta - 1 \nonumber \\
C &=& 6 \cos\theta \sin\theta \cos \Phi_B \nonumber \\
K &=& \left[ 3 \left( \cos \Theta_B \cos \theta - \sin\theta \sin\Theta_B \cos\Phi_B\right)^2-1 \right] \sin^2 \Theta_B.
\end{eqnarray}

The angles $(\theta_B,\phi_B)$ are obtained by doing the transformation from $(\Theta_B',\Phi_B+\pi)$ to the
vertical reference system.

\subsection{$\Phi_B' = \Phi_B+\pi/2$}
Under this change, we have:
\begin{eqnarray}
\cos 2\Phi_B' &=& -\cos 2\Phi_B, \\ \nonumber 
\sin 2\Phi_B' &=& -\sin 2\Phi_B, \\ \nonumber 
\cos \Phi_B' &=& -\sin \Phi_B, \\ \nonumber 
\sin \Phi_B' &=& \cos \Phi_B.
\end{eqnarray}
Following the same approach, we have to solve for $\Theta_B'$ in 
\begin{eqnarray}
&&\left[ 3 \left( \cos \Theta_B' \cos \theta + \sin\theta \sin\Theta_B' \sin\Phi_B\right)^2-1 \right] \sin^2 \Theta_B' \\ \nonumber 
&=&\left[ 3 \left( \cos \Theta_B \cos \theta - \sin\theta \sin\Theta_B \cos\Phi_B\right)^2-1 \right] \sin^2 \Theta_B.
\end{eqnarray}
The solution are obtained as the roots of the same equations as before but now
\begin{eqnarray}
A &=& -3\cos^2 \theta + 3\sin^2 \theta \sin^2 \Phi_B \nonumber \\
B &=& 3\cos^2 \theta - 1 \nonumber \\
C &=& 6 \cos\theta \sin\theta \sin \Phi_B \nonumber \\
K &=& -\left[ 3 \left( \cos \Theta_B \cos \theta - \sin\theta \sin\Theta_B \cos\Phi_B\right)^2-1 \right] \sin^2 \Theta_B.
\end{eqnarray}

The angles $(\theta_B,\phi_B)$ are obtained by doing the transformation from $(\Theta_B',\Phi_B+\pi/2)$ to the
vertical reference system.

\subsection{$\Phi_B' = \Phi_B-\pi/2$}
Under this change, we have:
\begin{eqnarray}
\cos 2\Phi_B' &=& -\cos 2\Phi_B, \\ \nonumber 
\sin 2\Phi_B' &=& -\sin 2\Phi_B, \\ \nonumber 
\cos \Phi_B' &=& \sin \Phi_B, \\ \nonumber 
\sin \Phi_B' &=& -\cos \Phi_B.
\end{eqnarray}
Following the same approach, we have to solve for $\Theta_B'$ in 
\begin{eqnarray}
&&\left[ 3 \left( \cos \Theta_B' \cos \theta + \sin\theta \sin\Theta_B' \sin\Phi_B\right)^2-1 \right] \sin^2 \Theta_B' \\ \nonumber 
&=&\left[ 3 \left( \cos \Theta_B \cos \theta - \sin\theta \sin\Theta_B \cos\Phi_B\right)^2-1 \right] \sin^2 \Theta_B.
\end{eqnarray}
The solution are obtained as the roots of the same equations as before but now
\begin{eqnarray}
A &=& -3\cos^2 \theta + 3\sin^2 \theta \sin^2 \Phi_B \nonumber \\
B &=& 3\cos^2 \theta - 1 \nonumber \\
C &=& -6 \cos\theta \sin\theta \sin \Phi_B \nonumber \\
K &=& -\left[ 3 \left( \cos \Theta_B \cos \theta - \sin\theta \sin\Theta_B \cos\Phi_B\right)^2-1 \right] \sin^2 \Theta_B.
\end{eqnarray}

The angles $(\theta_B,\phi_B)$ are obtained by doing the transformation from $(\Theta_B',\Phi_B-\pi/2)$ to the
vertical reference system.

\end{document}